\documentstyle[epsf,graphicx,11pt]{article}

\newcommand{\be}{\begin{equation}}
\newcommand{\ee}{\end{equation}}
\newcommand{\bea}{\begin{eqnarray}}
\newcommand{\eea}{\end{eqnarray}}
\newcommand{\bml}{\begin{mathletters}}
\newcommand{\eml}{\end{mathletters}}

\begin{document}






\title{Laplacians on lattices\\ }
\renewcommand{\thefootnote}{\fnsymbol{footnote}}
\author{Wojtek J. Zakrzewski\footnote{email: w.j.zakrzewski@durham.ac.uk}\\
Department of Mathematical Sciences,\\ University of Durham,\\
Durham DH1 3LE, UK}

\maketitle

\setlength{\footnotesep}{0.5\footnotesep}
\begin{abstract}
We consider some lattices and look at discrete Laplacians
on these lattices. In particular, we look at solutions
of the equation
$$\triangle(1) \phi = \triangle(2) Z, $$
where $\triangle(1)$ and $\triangle(2)$ denote two such Laplacians
on the same lattice.
We show that, in one dimension, when $\triangle(i)$ $i=1,2$ denote
$$\triangle(1) \phi = \phi(i+1)-\phi(i-1) -2\phi(i)$$
and
$$\triangle(2)Z=Z(i+
2)-Z(
i-2)-2Z(i)$$
this equation has a simple solution
$$\phi(i)=Z(i+1)+Z(i-1)+2Z(i).$$
We show that in two dimensions, when the system is considered on a 
hexagonal (honeycomb) lattice, we have a similar relation.
This is also true in three dimensions when we have a very special
lattice (tetrahedral with points inside).
We also briefly discuss how this relation generalises when we consider
other lattices.

\end{abstract}
\maketitle

\section{Introduction}

Recently, a Fr\"ohlich Hamitonian was studied on 
 two-dimensional, discrete, quadratic \cite{bpz1,bpz3,bpz2}, and
hexagonal lattices \cite{hz}. The resultant equations were rather
complicated to solve but when one restricted oneself to looking
at stationary fields, the equations simplified somewhat
and, in the case of hexagonal lattice, one could decouple them
and reduce them to the localised discrete nonlinear 
 Schr\"odinger
equation.

The reason for this decoupling lies in the observation
that, on a hexagonal lattice,
the equation
 \begin{equation}
\triangle(1) \phi \,= \,\lambda \,\triangle(2) Z
\label{our}
\end{equation}
has a simple localised solution for $Z$ as function of $\phi$.
Here $\triangle(i)$. $i=1,2$ refer to the two simplest discrete
Laplacians on a hexagonal lattice - i.e. those involving 4 and 7 lattice points 
(more details will be given later). 

A similar situation holds in one dimension and, in fact, was used
by Davydov \cite{davy} in his observation that the interaction of
an appropriate \hfil \break Schr\"odinger field, such as a field
describing amide I- vibration in biopolymers,
with the distortions of the underlying lattice,
results in the creation of a localised state which has, since then,
been refered to as Davydov's soliton.

In this paper, we look in detail at the eqution (\ref{our})
in the case of various lattices. 
We find that a regular lattice in one dimension, a hexagonal 
lattice in two dimensions and a tetrahedal lattice (with points
inside) in three dimensions are somewhat special
as it only in their cases the system possesses simple
localised solutions. 

We discuss the modifications that are required 
to have localised solutions for other (more complicated)
lattices.

\section{One dimension}

Consider a regular lattice as shown in Fig. 1.

\begin{figure}[htbp]
\unitlength1cm \hfil
\begin{picture}(10,1)
 \epsfxsize=9 cm \put(0,0){\epsffile{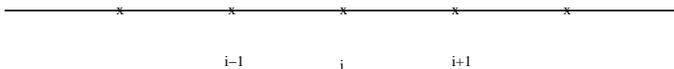}}
\end{picture}
\caption{\label{} 
One dimensional lattice.}
\end{figure}

\noindent For simplicity let us take the lattice spacing to be given 
by $a=1.$
Then
define $\triangle(1)P(i)$ as
\begin{equation}
\triangle(1)P(i)\,=\,P(i+1)+P(i-1)-2P(i).
\end{equation}
and
\begin{equation}
\triangle(2)
P(i)\,=\,P(i+2)
+P(i-2)
-2P(i)
\end{equation}
i.e. the same expression as above but with the displacement by two lattice 
units.

Then consider the equation:
\begin{equation}
\triangle(1) \phi(i) = \triangle(2) Z(i),
\label{oura}
\end{equation}
which we want to solve for $\phi(i)$
 in terms of $Z(k)$.

\subsection{Continuum limit}

Note that in the continuum limit our equation (\ref{oura})
reduces to
\begin{equation}
{\partial\sp2 \phi\over \partial x\sp2}\,
=\,
4 {\partial\sp2 Z\over
 \partial x\sp2},
 \end{equation}
  which clearly has a solution
  \begin{equation}
  \phi(x)\,=\,4Z(x)
  \end{equation}
  plus, of course a linear and
  a constant piece.
  
\subsection{Discrete case}
In this case it is easy to see that
 a solution is given by
 \begin{equation}
 \phi(i)\,=\,Z(i+1)+Z(i-1)+2Z(i).
 \label{sol}
 \end{equation}
 Of course, to this we can also add 
 $Ai+b$, which are the lattice equivalents
 of the linear and constant pieces
 of the continuum case.
 
 Note that our solution (\ref{sol}) has the correct continuum limit.
 Note also that its structure is very similar to the structure
 of $\triangle \phi(i)$ - except that, this time, all terms
 carry the $+$ sign.
 
 \section{Two dimensional Cases}
 \subsection{Square lattice}

\begin{figure}[htbp]
\unitlength1cm \hfil
\begin{picture}(10,6)
 \epsfxsize=7 cm \put(0,0){\epsffile{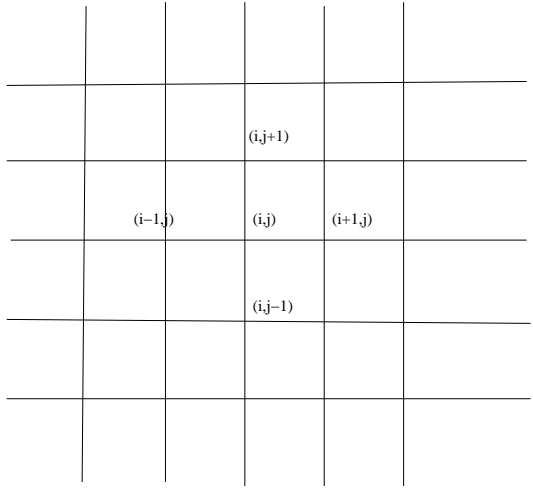}}
\end{picture}
\caption{\label{} 
Two-dimensional square lattice.} 
\end{figure}

 When we have a square lattice (see fig.2) we can consider various Laplacians.
 One of them, which we shall call $\triangle(1)$, involves
 the nearest lattice points and is defined as
 \begin{equation}
 \triangle(1) \phi(i,j
 )\,=\,\phi(i+1,j)+
 \phi(i-1,j)+\phi(i,j+1)+\phi(i,j-1)-4
\phi(i,j).
\end{equation}
The second 'obvious' Laplacian can be defined
as the one above with  shifts by two lattice points
ie
\begin{equation}
 \triangle(2) \phi(i,j
 )\,=\,\phi(i+2,j)+
 \phi(i-2,j)+\phi(i,j+2)+\phi(i,j-2)-4
\phi(i,j),
\end{equation}
or we could also 'involve' the corners (ie terms corresponding to $i\pm1$, $j\pm1$).

 Thus we could define $\triangle(2')\phi(i,j)$
 as

$$ \triangle(2') \phi(i,j)\,=\,\phi(i+2,j)+
 \phi(i-2,j)+\phi(i,j+2)+\phi(i,j-2)$$
  \begin{equation}\label{equ}
+\alpha\left(\phi(i-1,j-1)+\phi(i+1,j-1)
 +\phi(i-1,j+1)\right. 
 \end{equation}
 $$\left. +\phi(i-1,j-1)\right)
 -4(1+\alpha)
\phi(i,j),$$
for any value of $\alpha$.

The work mentioned in \cite{bpz1,bpz3,bpz2} required a solution of
\begin{equation}
\triangle(1)P(ij)\,=\,\mu\,\triangle(2) Z(ij),
\end{equation}
where $\triangle(2)$ is given by the expression as in (13).

Unfortunately, this equation has no simple local solution.
The problems is with the `corners'; the second Laplacian should involve
expressions at $i\pm1,j+\pm1$.

Thus it is possible to solve 
\begin{equation}
\triangle(1)P(ij)\,=\,\mu\,\triangle(2')Z(ij)
\end{equation}
for an appropriate choice of $\alpha$. To see this, in analogy with
(\ref{sol}) take
\begin{equation}
P(ij)\,=\,\lambda(Z(i-1,j)+Z(i+1,j)+Z(i,j-1)+Z(i,j+1)+4Z(i,j)).
\end{equation}
Then
\begin{equation}
\triangle(1)P(i,j)\,=\,\lambda\left(Z(i+1,j)+Z(i-2,j)+Z(i,j-2)+Z(i,j+2)\right.
\end{equation}
$$\left.+2Z(i+1,j+1)+2Z(i-1,j+1,+2Z(i+1,j-1)+2Z(i-1,j-1)\right),
$$
which is clearly a solution when $\alpha=2$ in (\ref{equ}). 
 
\subsection{hexagonal (honeycomb) lattice}

Next consider a hexagonal lattice as shown in Fig.3.

\begin{figure}[htbp]
\unitlength1cm \hfil
\begin{picture}(10,5)
 \epsfxsize=5 cm \put(0,0){\epsffile{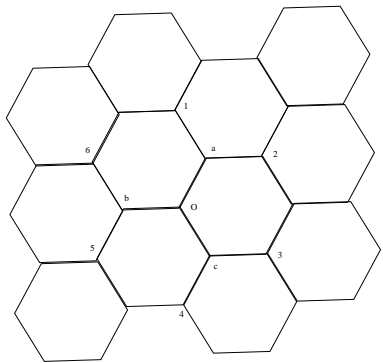}}
\end{picture}
\caption{\label{}Hexagonal (honeycomb) lattice.} 
\end{figure}

For $\triangle(1)$ take the 4-point Laplacian ( involving 0 and 3 `nearest' points); ie take
\begin{equation}
\triangle(1) P(0)\,=\, P(a)+P(b)+P(c)-3P(0)
\end{equation}

For $\triangle(2)$ take the Laplacian involving the 'next to the nearest' points - ie. the 7-point
Laplacian defined by
\begin{equation}
\triangle(2) Z(0)\,=\,Z(1)+Z(2)+Z(3)+Z(4)+Z(5)+Z(6)-6Z(0).
\end{equation}

Then consider
\begin{equation}
\label{eqq}
\triangle(1) P(0)\,=\,\mu\,\triangle(2) Z(0).
\end{equation}
In fact, this is the equation which arose in the study reported in \cite{hz}.
There it was shown that this equation has a simple local solution and that solution
is given by

\begin{equation}
P(0)\,=\,\mu(Z(a)+Z(b)+Z(c)+3Z(0)).
\end{equation}

Thus, in this sense, the hexagonal lattice resembles the most, the one dimensional
case; the double lattice spacing in the one diemensional case is replaced
by the double the angle between the directions on the lattice.

\subsection{triangular lattice}

Next we consider a triangular lattice as shown in Fig.4.

\begin{figure}[htbp]
\unitlength1cm \hfil
\begin{picture}(10,7)
 \epsfxsize=7 cm \put(0,0){\epsffile{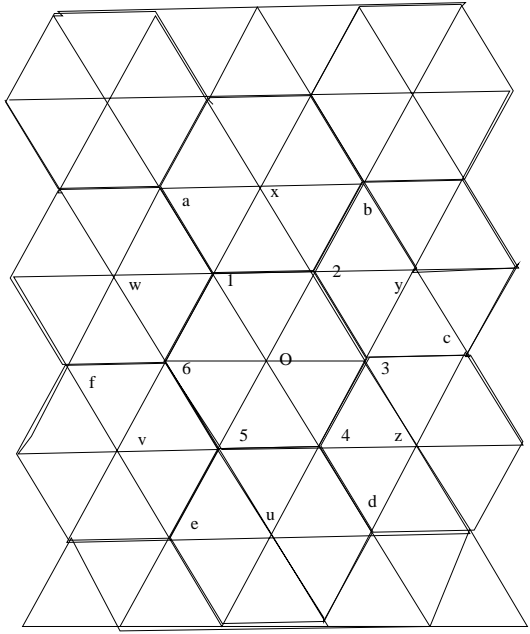}}
\end{picture}
\caption{\label{} The triangular lattice}
 \end{figure}

This time the obvious $\triangle(1)$ involves taking taking the 7-point 
Laplacian
\begin{equation}
\triangle(1)P(0)\,=\,P(1)+P(2)+P(3)+P(4)+P(5)+P(6)-6P(0).
\end{equation}

For the 'second' Laplacian we have more choice. We can take
\begin{equation}
\triangle(2)Z(0)\,=\,Z(a)+Z(b)+Z(c)+Z(d)+Z(e)+Z(f)-6Z(0).
\end{equation}
or we could add to it
\begin{equation}
+\,A(\,Z(u)+Z(v)+Z(w)+Z(x)+Z(y)+Z(z)-6Z(0)).
\end{equation}
However, in general, it is difficult to solve the equation
\begin{equation}
\triangle(1)P(0)=\triangle(2) Z(0)
\end{equation}
Only when $A=2$, i.e. for 
$\triangle(2')$ given by (24) and (25) with $A=2$, we have succeeded 
in finding a local solution.

This solution is given by
\begin{equation}
P(0)=Z(1)+Z(2)+Z(3)+Z(4)+Z(5)+Z(6)+4 Z(0).
\end{equation}

If guided by the experience gained from the previous cases we take
\begin{equation}
P(0)=Z(1)+Z(2)+Z(3)+Z(4)+Z(5)+Z(6)+6 Z(0).
\end{equation}
we find that the $Z$ field satisfies
\begin{equation}
Z(a)+Z(b)+Z(c)+Z(d)+Z(e)+Z(f) +2\left[ Z(1)+Z(2)+Z(3)+Z(4)\right.
\end{equation}
$$
\left. +Z(5)+Z(6)+Z(u)+Z(v)+Z(w)+Z(x)+Z(y)+Z(z)\right]\,-36Z(0)=0.$$

We see that in this case we have 19-point Laplacian - involving both the new
and the original lattice points.

One can consider other, more complicated, cases but it is clear that
the hexagonal lattice is very special as only for it, as in the one-dimensional
case for a regular lattice, we have a simple local solution of the problem (\ref{eqq}) with Laplacians
involving the `nearest' and `next-to-the-nearest' lattice points only.

\section{Three Dimensions}

The discussion of the previous sections generalises very easily to three dimensions.
First of all, for a regular cube-like lattice, for which in 2 dimensions
we had problems with `corner' terms, the situation is similar except that now we have more
such problems.

Thus if we take 
\begin{equation}
\triangle(1)P(i,j,k)=P(i+1,j,k)+P(i-1,j,k)+P(i,j-1,k)+P(i,j+1,k)
\end{equation}
$$+P(i,j,k-1)+P(i,j,k+1)-6P(i,j.k)
$$
and then consider
\begin{equation}
P(i,j,k)=Z(i+1,j,k)+Z(i-1,j,k)+Z(i,j-1,k)+Z(i,j+1,k)
\end{equation}
$$+Z(i,j,k-1)+Z(i,j,k+1)+6Z(i,j.k)
$$
we find that $Z(i,j,k)$ satisfies a rather complicated expression involving
the fields at all points $(i,j,k)$ which correspond to the shift of only one of
these indices by $\pm2$. There are 6 such terms and they all come with the coefficient
$=1$. In addition we also have to add all terms which involve the shift of two of the indices
$(i,j,k)$ by $\pm1$. There are 12 such terms and we have to add them all with the coefficient 
$=2$. And then finally we subtract $-24Z(0,0,0)$.
A little thought shows that this is the obvious generalisation of the square lattice
which we discussed in the previous section.

However, 
we have a regular lattice which is a `natural' 3-dimensional generalisation of
the hexagonal lattice in 2 dimensions.
Such lattice can be constructed in the following way:

\begin{figure}[htbp]
\unitlength1cm \hfil
\begin{picture}(10,7)
 \epsfxsize=7 cm \put(0,0){\epsffile{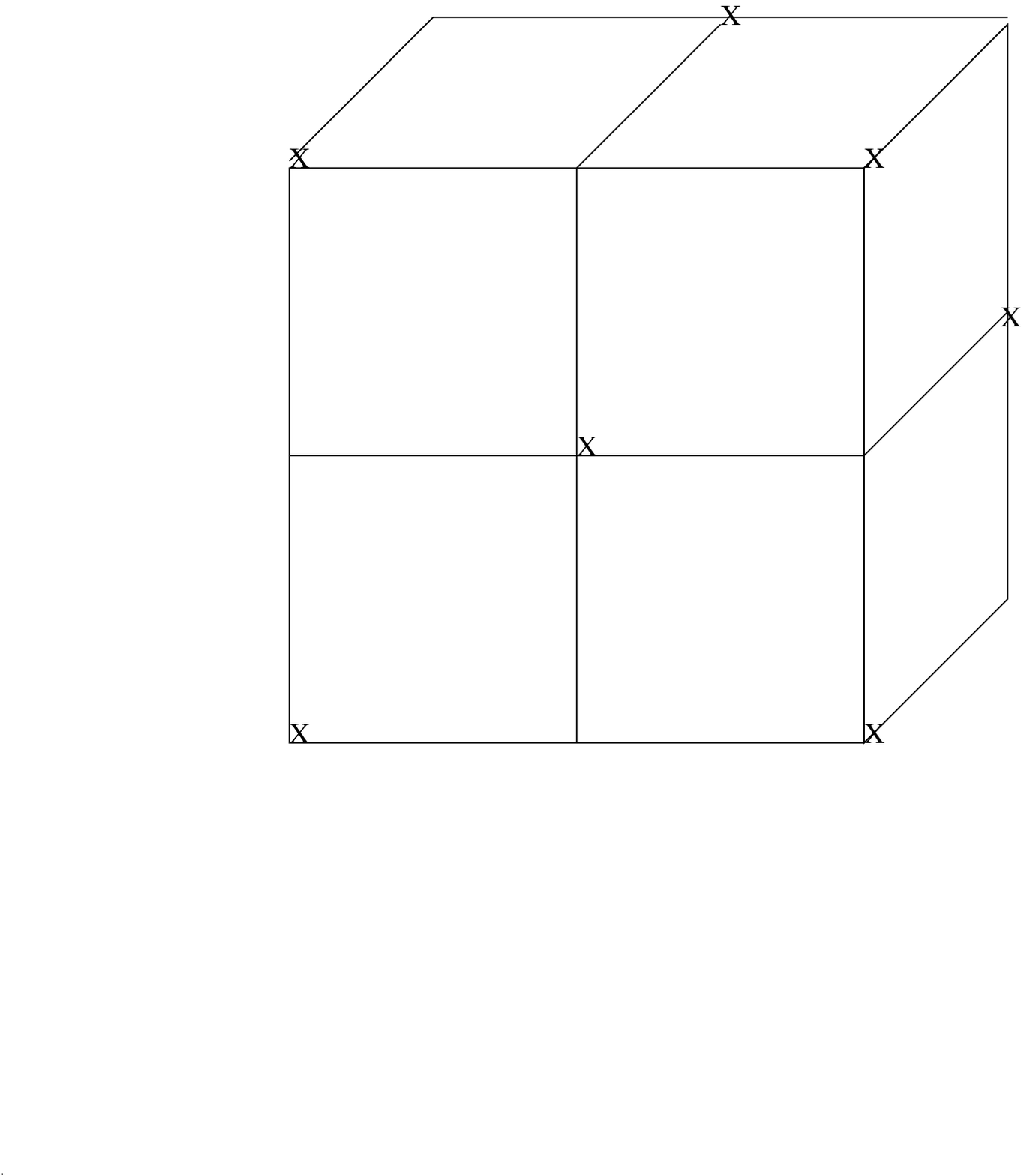}}
\end{picture}
\caption{\label{} The three-dimensional lattice. Points $A$ shown.
In the centre of each other cube there is a point $B$. Points $B$, togetherm
form another lattice - like the one shown but displaced by ${1\over 2}$
lattice spacing in all directions.} 
\end{figure}

Take a regular cube lattice and delete every other lattice point as indicated in Fig.5.
Then  place a similar lattice at the points which are the centres of the 
cubes of the previous lattice. This way we have a body centred lattice with many 
points missing.

Consider then the points of the original cube lattice as points of type $A$ and those
of the additional lattice as points of type $B$. Then the nearest neighbours
of each point $A$ are 4 lattice points of type $B$ and vice-versa.
The next to the nearest neigbours (ntn), however, come from the original lattice;
ie for a point type $A$ - are points of type $A$, and for the point $B$ are
points of type $B$. Each point $A$ has 4 nearest neighbours and 12 ntn points.
And, by construction, this is the case for both types of lattice points.

Clearly we have two obvious Laplacians that we can consider - those involving
the nearest neigbours and those involving only the ntns.

Let us consider, for simplicity,  the point $B$ located at (0,0,0)
inside a cube of size $2a$, with points, of type $A$ located at (1,-1,1), (-1,-1, -1)
and (1,1,-1) and (-1,1,1,) (we are using the convention of $x$ - horizontally to the right,
$z$ vertically up and $y$ away from the observer).
Then we can define

\begin{equation}
\triangle(1)P(0,0,0)\,=\,P(1,-1,1)+P(-1,-1,-1)+P(1,1,-1)\end{equation}
$$+P(-1,1,1)-4P(0,0,0).$$

In the continuum limit
this expression, clearly, reduces to the Laplacian of $P$ ie
 $(\partial\sp2_x +  \partial\sp2_y 
+ \partial\sp2_z)P$.

For the second Laplacian we take ntn points. Notice that they correspond
to points $\pm2$ and $0$ such that one of the coordinates
of $x$, $y$ and $z$ = 0 while the others take values $\pm2$.
 Thus we define
$$
\triangle(2)P(0,0,0)\,=\,P(-2,-2,0)+P(2,-2,0)+P(-2,2,0)+P(2,2,0)$$
\begin{equation}+P(-2,0,-2)+P(-2,0,2)+P(2,0,-2)+P(2,0,2)\end{equation}
$$+P(-2,-2,0)+P(-2,2,0)+P(2,-2,0)+P(2,2,0).$$

Now we consider our equation:
\begin{equation}
\label{EQ}
\triangle(1)P\,=\, \alpha \triangle(2) Z,
\end{equation}
where $P$ and $Z$ are taken at the same lattice point, say $(0,0,0)$.

Guided by our experience from the lower dimensional cases we 
take
\begin{equation}
P(0,0,0)=\beta (Z(1,-1,1)+Z(-1,-1,-1)+Z(1,1,-1)\end{equation}
$$+Z(-1,1,1)+4Z(0,0,0)).$$

Then, it is a matter of simple algebra to check that $P$ does indeed satisfy
(\ref{EQ}) when $\beta=\alpha$.
A little thought shows that our lattice is really tetrahedral in nature (with points
inside it), hence it is a natural generalisation of the hexagonal lattice in 2 dimensions.

\section{Conclusions}

In this short note we have discussed various lattices and Laplacians defined on these 
lattices. Our main interest was the relation between various Laplacians
and the existence or not of fully localiseed
solutions of the equation
\begin{equation}
\triangle(1)P(0)=\triangle(2) Z(0)
\end{equation} involving these Laplacians.

We have found that, in one dimension,  this equation a localised solution 
when the lattice is regular and $\triangle(1)$ and $\triangle(2)$ involve 
the Laplacians constructed with the `nearest' and `next-to-the-nearest' 
lattice points. Moreover the points in each Laplacian are equidistant 
from the central point.

These conditions are also required in higher dimensional lattices where they
are much more stringent. Thus in two dimensions they require that the lattice 
is hexagonal (honeycomb) and then $\triangle(1)$ and $\triangle(2)$ involve
1+3 point and 1+6 point Laplacians. For other lattices the Laplacians 
mix points of different distance from the central point and we have not succeeded 
in finding a localised solution of (15).

In three dimensions the relevant lattice is the `body centred tetrahedral lattice'. In this case 
$\triangle(1)$ involves 1+4 lattice points and $\triangle(2)$ - 1+12 points.

Note that in 1 dimension - our Laplacians involve doubling of distance,
in 2 dim - doubling the angle, and in 3 dim - doubling the spherical angles.

Noting the pattern of our results we expect that our results generalise to higher
dimensions, where in $D$ dimensions we expect $\triangle(1)$ to involve $1+(D+1)$ points
and $\triangle(2)$ $1+D(D+1)$ points.
But lattices with such Laplacians do not appear to be physical so we have not attempted to 
construct them.

\vskip 1cm

{\bf Acknowledgements.} {We would like to thank H. B\"uttner, L. Brizhik, A. Eremko, B. Hartmann, B. Piette
and C. Mueller for interest, helpful comments and suggestions.
This work was completed when the author was visiting the Physics Institute
at the University of Bayreuth. We wish to thank the University of Bayreuth
for its hospitality and DAAD for the award of a grant that made our visit 
to Bayreuth possible.}

\end{document}